# On the Benefits of Bandwidth Limiting in Decentralized Vector Multiple Access Channels

Samir M. Perlaza[1], Mérouane Debbah[2], Samson Lasaulce[3], Hanna Bogucka[4]

[1] France Telecom R&D - Orange Labs Paris. France
Samir.MedinaPerlaza@orange-ftgroup.com

[2] Alcatel Lucent Chair in Flexible Radio - SUPELEC. France
Merouane.Debbah@supelec.fr

[3] Laboratoire des Signaux et Systèmes (LSS) - CNRS, SUPELEC, Univ. Paris Sud. France
Samson.Lasaulce@lss.supelec.fr

[4] Poznan University of Technology. Poland
hbogucka@et.put.poznan.pl

*Abstract*—We study the network spectral efficiency of decentralized vector multiple access channels (MACs) when the number of accessible dimensions per transmitter is strategically limited. Considering each dimension as a frequency band, we call this limiting process bandwidth limiting (BL). Assuming that each transmitter maximizes its own data rate by water-filling over the available frequency bands, we consider two scenarios. In the first scenario, transmitters use non-intersecting sets of bands (spectral resource partition), and in the second one, they freely exploit all the available frequency bands (spectral resource sharing). In the latter case, successive interference cancelation (SIC) is used. We show the existence of an optimal number of dimensions that a transmitter must use in order to maximize the network performance measured in terms of spectral efficiency. We provide a closed form expression for the optimal number of accessible bands in the first scenario. Such an optimum point, depends on the number of active transmitters, the number of available frequency bands and the different signal-to-noise ratios. In the second scenario, we show that BL does not bring a significant improvement on the network spectral efficiency, when all transmitters use the same BL policy. For both scenarios, we provide simulation results to validate our conclusions.

## I. INTRODUCTION

In a vector multiple access channel (MAC), a large set of transmitters share a limited set of frequency bands (channels) to communicate with a unique receiver [1]. When there exists a central controller (normally the receiver) a capacity achieving power allocation can be implemented by using an iterative water-filling algorithm [2], [3]. In this case, the central controller knows the transmission parameters and instantaneous channel realizations of each transmitter over each channel. Thus, it is able to solve the global optimization problem and feed back the optimal power levels to each transmitter. However, in the absence of a central controller or the impossibility to interchange signaling messages between the transmitters to obtain a complete information of the network, achieving capacity becomes a non-trivial task [4]. Here, game theory has played a remarkable role, but solutions remain being highly suboptimal due to the lack of global information [5].

To overcome this sub optimality, imposing orthogonal communications between transmitters using only a single channel has been a well-accepted solution, e.g., IEEE802.11 networks.

In this case, transmitters reduce the mutual interference and only overcome the interference of transmitters sharing the same channel. Nonetheless, up to the knowledge of the authors, the choices of the total number of available channels [6] as well as limiting the bandwidth to a single channel have been done in an ad hoc manner. This paper provides an analysis of the benefits of bandwidth limiting (BL), i.e., reducing the number of channels each transmitter can use in vector MAC. More specifically, we provide an answer to the following question: is it worth to limit the number of channels each transmitter might use regarding the network spectral efficiency?

We consider two scenarios. In the first scenario, transmitters have to use non-intersecting sets of channels. In the second one, transmitters can freely exploit all the available channels. In the second case, the receiver implements multiuser decoding and successive interference cancelation (SIC). Here, each transmitter is aware of both the decoding order and its respective noise plus interference levels. Under these conditions, the optimal decentralized policy for each transmitter to maximize its own data rate is to use a water-filling power allocation scheme considering the multiple access interference as noise [2].

Our work is motivated by the following reasoning. In the first scenario, the fact that a transmitter uses several channels significantly reduces the total number of active transmitters. For instance, in the high signal to noise ratio (SNR) regime and considering a finite set of channels, few transmitters might occupy all the available channels. Following a water-filling power allocation, a given transmitter allocates the highest power levels to the channels with the highest gains. Then, the channels being used with low powers might have a negligible impact on its individual data rate. However, no other transmitter can access the spectrum, even though, a higher rate can be obtained by another transmitter. In the second scenario, having a transmitter using several channels does not reduce the number of active transmitters since they can co-exist in the same channels. However, the multiple access interference produced by the transmitters decoded in the last places might significantly reduce the data rates of those decoded in the first places. More specifically, the gain in data rate obtained by transmitters decoded in last places on certain channels, might not compensate the loss of the transmitters decoded in first places.

In both cases, this effect stems from the fact that transmitters maximize its own spectral efficiency independently of the others. This resulting suboptimal usage of the spectrum recall us the dilemma presented in [7], known as *the tragedy of the commons*. Therein, it is shown how multiple individuals acting independently in their own self-interest can ultimately destroy a shared limited resource even when it is clearly not in anyone's long term interest. One of the solutions for these dilemma is to introduce regulation by an authority. In this paper, we analyze this point of view and try to find the rules (in terms of BL) each transmitter must follow to maximize the network spectral efficiency.

We show the existence of an optimal BL point in the first scenario. Such optimal number of channels is a function of the total number of active transmitters, the total number of channels, and the different signal to noise ratios (SNR). We present simulations where we observe a significant gain in terms of spectral efficiency. In the second scenario, we show that BL does not bring a significant improvement if all transmitters use the same BL parameter, i.e., all transmitters are limited to use the same number of channels.

## II. SYSTEM MODEL

Consider a set $\mathcal{K} = \{1, \ldots, K\}$ of transmitters communicating with a unique receiver using a set $\mathcal{N} = \{1, \ldots, N\}$ of equally spaced frequency bands (channels) as shown in Fig. 1. In the information theory jargon, this network topology is known as vector MAC or parallel MAC [1]. Transmitters arrive sequentially to the network. Their index in the set $\mathcal{K}$ shows the order of arrival. All the radio devices are equipped with a unique antenna. Transmitter $k \in \mathcal{K}$ is able to simultaneously transmit over all the channels subject to a power-limitation,

$$\forall k \in \mathcal{K}, \quad \frac{1}{N}\sum_{n=1}^{N} p_{k,n} \leqslant p_{k,\max}, \quad (1)$$

where $p_{k,n}$ and $Np_{k,\max}$ denote the transmit power over channel $n$ and the maximum transmittable power of transmitter $k$. Without any loss of generality, we assume that all transmitters are limited by the same maximum transmittable power level, i.e., $\forall k \in \mathcal{K}$ and $\forall n \in \mathcal{N}$, $p_{k,\max} = p_{\max}$.

We denote the channel coefficients in the frequency domain between the receiver and transmitter $k$ over channel $n$ by $h_{k,n}$. We assume that for the whole transmission duration, all the channel realizations remain constant. For all $n \in \mathcal{N}$ and for all $k \in \mathcal{K}$, $h_{k,n}$ is a realization of a complex random variable $h$ with independent and identically distributed (i.i.d) Gaussian real and imaginary parts with zero mean and variance $\frac{1}{2}$. The channel gain is denoted by $g_{k,n} = ||h_{k,n}||^2$. Then, the channel gains can be modeled by realizations of a random variable $g$ with exponential distributions with parameter $\rho = 1$, whose cumulative distribution function (c.d.f) and probability density function (p.d.f) are denoted by $F_g(\lambda) = 1 - e^{-\lambda}$ and $f_g(\lambda) = e^{-\lambda}$, respectively. The received signals sampled at symbol rate can be written as a vector $\boldsymbol{y} = (y_1, \ldots, y_N)$ where the entries $y_n$ for all $n \in \mathcal{N}$ represent the received signal over channel $n$. Hence,

$$\boldsymbol{y} = \sum_{k=1}^{K} \boldsymbol{H}_k \boldsymbol{s}_k + \boldsymbol{w}, \quad (2)$$

where $\boldsymbol{H}_k$ is an $N$-dimensional diagonal matrix with main diagonal $(h_{k,1}, \ldots, h_{k,N})$. The $N$-dimensional vector $\boldsymbol{s}_k = (s_{k,1}, \ldots, s_{k,N})$ represents the symbols transmitted by transmitter $k$ over each channel. The power allocation profile of transmitter $k$, the vector $(p_{k,1}, \ldots, p_{k,N})$, is the diagonal of the diagonal matrix $\boldsymbol{P}_k = \mathbb{E}\left[\boldsymbol{s}_k \boldsymbol{s}_k^H\right]$. The $N$-dimensional vector $\boldsymbol{w}$ represents the noise at the receiver. Its entries, $w_n$ for all $n \in \mathcal{N}$, are modeled by a complex circularly symmetric additive white Gaussian noise (AWGN) process with zero mean and variance $\sigma^2$.

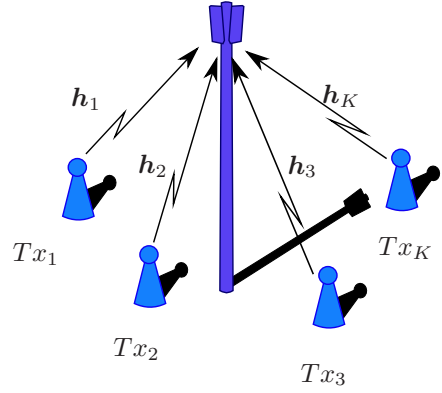

Fig. 1. Vector multiple access channel with $K$ transmitters and $N$ available channels with $\boldsymbol{h_k} = (h_{k,1}, \ldots, h_{k,N})$ for all $k \in \mathcal{K}$.

Regarding the channel state information (CSI) we assume that each transmitter perfectly knows its own channel coefficients and the noise plus interference level at each channel. This is the case when transmitters are able to sense its environment or the receiver feeds back this parameter as a signaling message to all the transmitters.

We denote the set of channels being used by transmitter $k$ by $\mathcal{L}_k$, i.e., $\forall k \in \mathcal{K}$ and $\forall n \in \mathcal{L}_k$, $p_{k,n} \neq 0$, and $\forall m \in \mathcal{N} \setminus \mathcal{L}_k$, $p_{k,m} = 0$. Depending on the conditions over each set $\mathcal{L}_k$, for all $k \in \mathcal{K}$, we consider two different scenarios.

### A. Scenario 1: Spectral Resource partition

In this scenario, a given channel cannot be used by more than one transmitter. Thus, this is equivalent to define the sets $\mathcal{L}_k$ for all $k \in \mathcal{K}$ as a partition of the set $\mathcal{N}$, i.e.,

- $\forall (j,k) \in \mathcal{K}^2$ and $j \neq k$, $\mathcal{L}_j \cap \mathcal{L}_k = \emptyset$,
- $\forall (j,k) \in \mathcal{K}^2$ and $j \neq k$, $\mathcal{L}_j \cup \mathcal{L}_k \subseteq \mathcal{N}$,
- $\forall k \in \mathcal{K}$, $|\mathcal{L}_k| > 0$.

Due to the asynchronous arrival of the users, we assume that there exists an order to access the set of channels $\mathcal{N}$. We index the transmitters such that transmitter $k \in \mathcal{K}$ arrives in the $k^{th}$ position.

## B. Scenario 2: Spectral Resource Sharing

In this scenario we allow several transmitters to use the same channels. Thus, this is equivalent to define the sets $\mathcal{L}_k$ for all $k \in \mathcal{K}$ as a cover of the set $\mathcal{N}$, i.e.,

- $\forall k \in \mathcal{K}, \quad \mathcal{L}_k \subseteq \mathcal{N}$,
- $\forall k \in \mathcal{K}, |\mathcal{L}_k| > 0$.

We assume that the receiver performs multiuser decoding and successive interference cancelation (SIC) at each channel. The decoding order is the same in all channels and it is related to the arrival order. Here, transmitter $k \in \mathcal{K}$, arriving in the $k^{th}$ position, is decoded in the $K - k + 1$ position.

To make a difference between scenario 1 and 2, we use the super index $^{(i)}$ with $i \in \{1, 2\}$ for all the sets and variables associated with each of them. The noise plus multiple access interference (MAI) for transmitter $k$ over channel $n$ is denoted by $\alpha_{k,n}^{(i)}$, where $\alpha_{k,n}^{(1)} = \sigma^2$ and $\alpha_{k,n}^{(2)} = \sigma^2 + \sum_{j=1}^{k-1} p_{j,n} g_{j,n}$, where, for all $n \in \mathcal{N}$, $p_{0,n} = 0$ and $g_{0,n} = 0$. The SINR for transmitter $k$ over channel $n$ is denoted by $\gamma_{k,n}^{(i)}$ and $\forall k \in \mathcal{K}$ and $\forall n \in \mathcal{N}$,

$$\gamma_{k,n}^{(i)} = \frac{p_{k,n} g_{k,n}}{\alpha_{k,n}^{(i)}}. \quad (3)$$

In both scenarios each transmitter $k \in \mathcal{K}$ aims to maximize its own data rate $R_k(\boldsymbol{\gamma}_k)^{(i)}$, i.e.,

$$R_k(\boldsymbol{\gamma}_k)^{(i)} = \sum_{n=1}^{N} \log_2\left(1 + \gamma_{k,n}^{(i)}\right), \quad (4)$$

with $\boldsymbol{\gamma}_k^{(i)} = (\gamma_{k,1}^{(i)}, \ldots, \gamma_{k,N}^{(i)})$ subject to its power limitations and independently of the data rate of the other devices. We explain this process in the next section.

## III. INDIVIDUAL SPECTRAL EFFICIENCY

Assuming that each channel bandwidth is normalized to 1 Hz, a given transmitter $k$ sets out its transmit power levels $p_{k,n}$, $\forall n \in \mathcal{N}$ by solving the optimization problem (OP)

$$\begin{aligned} \max_{p_{k,n}, \forall n \in \mathcal{Z}_k^{(i)}} & \sum_{n \in \mathcal{Z}_k^{(i)}} \log_2\left(1 + \gamma_{k,n}^{(i)}\right) \\ \text{s.t.} & \frac{1}{N} \sum_{n \in \mathcal{Z}_k^{(i)}} p_{k,n} \leqslant p_{\max}, \end{aligned} \quad (5)$$

where, for all $k \in \mathcal{K}$, the set $\mathcal{Z}_k^{(1)} = \mathcal{N} \setminus \mathcal{L}_1 \cup \ldots \cup \mathcal{L}_{k-1}$, with $\mathcal{L}_0 = \emptyset$, and $\mathcal{Z}_k^{(2)} = \mathcal{N}$. Thus, $\mathcal{Z}_k^{(i)}$ is the set of channels available for user $k$ in scenario $i$.

The solution to the OP in (Eq. 5) is given in [1] and thus, we only provide the solution hereafter; $\forall k \in \mathcal{K}$ and $\forall n \in \mathcal{Z}_k^{(i)}$,

$$p_{k,n} = \left[\beta - \frac{\alpha_{k,n}^{(i)}}{g_{k,n}}\right]^+, \quad (6)$$

and, $\forall k \in \mathcal{K}$ and $\forall n \in \mathcal{N} \setminus \mathcal{Z}_k^{(1)}$,

$$p_{k,n} = 0. \quad (7)$$

Here, the operator $[.]^+$ is the same as $\max(0, .)$. Given the sets $\mathcal{A}$, $\mathcal{B}$ and the complement of the latter, $\mathcal{B}'$, in a given universal set, the operation $\mathcal{A} \setminus \mathcal{B} = \mathcal{A} \cap \mathcal{B}'$. The term $\beta$ is a Lagrangian multiplier, known as water-level, chosen to satisfy (Eq. 1). The transmit power levels in (Eq. 6) can be iteratively obtained by using the water-filling algorithm described in [1]. From expression (Eq. 6), it can be implied that $\mathcal{L}_k^{(i)} \subseteq \mathcal{Z}_k^{(i)}$.

Once the OP in (Eq. 5) has been solved, the data rate per channel of transmitter $k \in \mathcal{K}$, is

$$\bar{R}_k(\boldsymbol{\gamma}_k)^{(i)} = \frac{1}{|\mathcal{Z}_k^{(i)}|} \sum_{n \in \mathcal{Z}_k^{(i)}} \log_2\left(1 + \gamma_{k,n}^{(i)}\right), \quad (8)$$

and then, its spectral efficiency $\Phi_k$ is

$$\Phi_k^{(i)} = \underbrace{\frac{|\mathcal{Z}_k^{(i)}|}{N}}_{\Omega_k^{(i)}} \bar{R}_k(\boldsymbol{\gamma}_k)^{(i)}, \quad (9)$$

where, $\Omega_k^{(i)}$ represents the fraction of spectrum accessible for transmitter $k$. Note that due to the decentralized nature of the network, the individual spectral efficiency is maximized independently by each transmitter. As described in [7], it might lead to significant losses in the network spectral efficiency. We study this effect in the next section.

## IV. NETWORK SPECTRAL EFFICIENCY

We define the network spectral efficiency (NSE) $\Phi^{(i)}$ as

$$\begin{aligned} \Phi^{(i)} &= \sum_{k=1}^{K} \Phi_k^{(i)} \\ \Phi^{(i)} &= \sum_{k=1}^{K} \Omega_k^{(i)} \bar{R}_k^{(i)}, \end{aligned} \quad (10)$$

for both scenarios, spectral resource partition (scenario 1) and spectral resource sharing (scenario 2). We analyze the NSE in the asymptotic regime, i.e., we assume that both the number of channels $(N)$ and the number of transmitters $(K)$ grow to infinity at a constant ratio $\frac{N}{K} = \alpha < \infty$. Under these conditions, we determine the NSE in the absence of bandwidth limiting and we provide closed form expressions in both cases. Later, we determine the NSE using BL and provide closed form expressions.

### A. NSE Without BL

A first result on the analysis of NSE in the absence of BL for the case of spectral resource partition is presented in [8], [9]. We revisit those results and extend it to the case of resource sharing.

*1) Scenario 1: Spectral Resource Partition:* Following the same line of the analysis presented in [8], we have that in the asymptotic regime the data rate per channel for a given transmitter $k$ is

$$\bar{R}_k(\boldsymbol{\gamma}_k)^{(1)} \xrightarrow{N \to \infty} \underbrace{\int_0^\infty \log_2\left(1 + \frac{p_k(\lambda)\lambda}{\sigma^2}\right) \mathrm{d}F_g(\lambda)}_{\bar{R}_{k,\infty}^{(1)}} \quad (11)$$

where the functions $p_k(\lambda)$ for all $k \in \mathcal{K}$, satisfy the power constraints,

$$\int_0^\infty p_k(\lambda) \mathrm{d}F_g(\lambda) = p_{\max}. \quad (12)$$

The function $p_k(\lambda)$, $\forall k \in \mathcal{K}$, which maximizes expression (Eq. 11) subject to expression (Eq. 12) is also a water-filling solution, i.e.,

$$p_k(\lambda) = \left[\beta_k - \frac{\sigma^2}{\lambda}\right]^+. \quad (13)$$

Note that since all the channel coefficients are drawn from the same probability distribution $f_g(\lambda)$ described in Sec. II and all the transmitters have the same maximum transmittable power level, we can write that $\forall k \in \mathcal{K}$, $\bar{R}_{k,\infty}^{(1)} = \bar{R}_\infty^{(1)}$. Hence, the water-level $\beta_k$ satisfying the condition (Eq. 13) is the same for all the transmitters. By combining expression (Eq. 13) and (Eq. 12), we obtain the water-level $\beta_k = \beta^*$, $\forall k \in \mathcal{K}$ in the asymptotic regime by solving the equation

$$\int_{\frac{\sigma^2}{\beta^*}}^{\infty} \left(\beta^* - \frac{\sigma^2}{\lambda}\right) dF_g(\lambda) - p_{\max} = 0. \quad (14)$$

The fraction $\Omega_k^{(1)}$, for all $k \in \mathcal{K}$, can be approximated in the asymptotic regime by $\Omega_{k,\infty}^{(1)}$ [8]

$$\Omega_{k,\infty}^{(1)} \stackrel{N \to \infty}{\longrightarrow} \left(\Omega_\infty^{(1)}\right)^{k-1}$$

where,

$$\Omega_\infty^{(1)} = \Pr\left(\beta^* < \frac{\sigma^2}{\lambda}\right) = \int_0^{\frac{\sigma^2}{\beta^*}} dF_g(\lambda) \leq 1. \quad (15)$$

Then, the NSE (Eq. 10) in the asymptotic regime $\Phi_\infty^{(1)}$ is

$$\Phi_\infty^{(1)} = \sum_{i=1}^{K} \left(\Omega_{k,\infty}^{(1)}\right)^{i-1} \bar{R}_\infty^{(1)} = \frac{1 - \left(\Omega_\infty^{(1)}\right)^K}{1 - \Omega_\infty^{(1)}} \bar{R}_\infty^{(1)}. \quad (16)$$

*2) Scenario 2: Spectral Resource Sharing:* In the asymptotic regime, we can approximate the data rate per channel of transmitter $k \in \mathcal{K}$ as

$$\bar{R}_k(\boldsymbol{\gamma}_k)^{(2)} \stackrel{N \to \infty}{\longrightarrow} \underbrace{\int_0^\infty \cdots \int_0^\infty \Gamma_k'(\boldsymbol{\lambda}_k) dF_g(\lambda_k)..dF_g(\lambda_1)}_{\bar{R}_{k,\infty}^{(2)}} \quad (17)$$

with $\boldsymbol{\lambda}_k = (\lambda_1, \ldots, \lambda_k)$ and

$$\Gamma_k'(\boldsymbol{\lambda}_k) = \log_2\left(1 + \frac{p_k(\lambda_k)\lambda_k}{\sigma^2 + \sum_{j=1}^{k-1} p_j(\lambda_j)\lambda_j}\right), \quad (18)$$

where $p_0(\lambda_0) = 0$ for all $\lambda_0 \in \mathbb{R}$. Additionally, for all $k \in \mathcal{K}$, the functions $p_k(\lambda_k)$ satisfy the power constraints shown in (Eq. 12). As shown in the previous section, the maximization of (Eq. 17) is a water-filling solution. For transmitter $k = 1$, it yields expression (Eq. 13) and for all $1 < k \leq K$,

$$p_k(\boldsymbol{\lambda}_k) = \left[\beta_k^* - \frac{\beta_{k-1}^* \lambda_{k-1}}{\lambda_k}\right]^2. \quad (19)$$

Now, by plugging expression (Eq. 19) in (Eq. 17), we obtain that for all $1 < k \leq K$,

$$\bar{R}_{k,\infty}^{(2)} = \int_{\frac{\sigma^2}{\beta_1^*}}^{\infty} \int_{\frac{\beta_1^* \lambda_1}{\beta_2^*}}^{\infty} \cdots \int_{\frac{\beta_{k-1}^* \lambda_{k-1}}{\beta_k^*}}^{\infty} \Gamma_k(\boldsymbol{\lambda}_k) dF_g(\lambda_k)...dF_g(\lambda_1) \quad (20)$$

with $\Gamma_k(\boldsymbol{\lambda}_k) = \log_2\left(\frac{\beta_k^* \lambda_k}{\beta_{k-1}^* \lambda_{k-1}}\right)$, whereas for transmitter $k = 1$, the asymptotic data rate is given by expression (Eq. 11). The water-levels $\beta_k^*$ in (Eq. 20) for all $k \in \mathcal{K}$ are the solution to expression (Eq. 14) in the case $k = 1$, and

$$p_{\max} = \int_{\frac{\sigma^2}{\beta_1^*}}^{\infty} \int_{\frac{\beta_1^* \lambda_1}{\beta_2^*}}^{\infty} \cdots \int_{\frac{\beta_{k-1}^* \lambda_{k-1}}{\beta_k^*}}^{\infty} p_k(\boldsymbol{\lambda}_k) dF_g(\lambda_k)...dF_g(\lambda_1), \quad (21)$$

in the case $k \in \mathcal{K} \setminus \{1\}$. Thus, the spectral efficiency of transmitter $k \in \mathcal{K}$ is equivalent to its data rate per channel, i.e.,

$$\Phi_{k,\infty}^{(2)} = \bar{R}_{k,\infty}^{(2)}. \quad (22)$$

Here, the factor $\Omega_k^{(2)}$ in (Eq. 10), is $\Omega_k^{(2)} = 1$ for all $k \in \mathcal{K}$. This is because each transmitter can access all the channels regardless of its order of arrival.

By developing expression (Eq. 21) and (Eq. 20) using the p.d.f of the channel gains $f_g(\lambda)$ described in Sec. II, we arrive to the following conclusions,

$$\forall k \in \mathcal{K} \setminus \{1\}, \quad \beta_{k-1}^* \leq \beta_k^*, \quad (23)$$

and

$$\forall k \in \mathcal{K} \setminus \{1\}, \quad \bar{R}_{k-1,\infty}^{(2)} \geq \bar{R}_{k,\infty}^{(2)}, \quad (24)$$

respectively. Then, the NSE (Eq. 10) in the asymptotic regime $\Phi_\infty^{(2)}$ is

$$\Phi_\infty^{(2)} = \sum_{k=1}^{K} \bar{R}_{k,\infty}^{(2)}. \quad (25)$$

In Sec. VI, we compare both asymptotic and non-asymptotic expressions to validate our statements.

### B. NSE With BL

Now, we limit the number of channels each transmitter can use. For the ease of calculations, we keep the conditions that both $K$ and $N$ grow to infinity at the same rate, i.e., $N \to \infty$, and $K \to \infty$, and $\frac{N}{K} = \alpha < \infty$. To provide at least one channel to each user, we assume that $\alpha \geq 1$.

*1) Scenario 1: Spectral Resource Partition:* When the number of accessible channels for transmitter $k \in \mathcal{K}$ is limited to $L \in \mathcal{N}$ channels, the fraction of accessible spectrum $\Omega_k^{(1,\text{BL})}$ for each transmitter is

$$\Omega_{k,\infty}^{(1,\text{BL})} = \min\left\{\Pr\left(\beta^* < \frac{\sigma^2}{\lambda}\right), \frac{L}{N}\right\}. \quad (26)$$

Then, BL has an effect if and only if $\frac{L}{N} < \Pr\left(\beta^* \leq \frac{\sigma^2}{\lambda}\right)$. This condition is equivalent to state that we should limit the transmitters to use a smaller number of channels of that used on the absence of BL. Hence, under the asymptotical assumptions, we have that

$$\Omega_{k,\infty}^{(1,\text{BL})} = \frac{L}{N}, \quad (27)$$

and

$$\Phi_\infty^{(1,\text{BL})} = \sum_{i=1}^{K} \Omega_{k,\infty}^{(1,\text{BL})} \bar{R}_\infty = \frac{KL}{N} \bar{R}_\infty. \quad (28)$$

*2) Scenario 2: Spectral Resource Sharing:* In this scenario, each transmitter can access all the channels and thus, $\Omega_{k,\infty}^{(2)} = 1$. When we limit the bandwidth for transmitter $k$ we impose that $\Omega_{k,\infty}^{(2),\text{BL}} \leqslant \Omega_{k,\infty}^{(2)}$. Then, the NSE under BL is

$$\Phi_{\infty}^{(2),\text{BL}} = \sum_{k=1}^{K} \Omega_{k,\infty}^{(2),\text{BL}} \bar{R}_{k,\infty}^{(2),\text{BL}}. \qquad (29)$$

We have provided expressions for the NSE in both absence and presence of BL in the asymptotic regime. Now, it remains to determine the conditions over which BL brings benefits to the network in terms of spectral efficiency.

## V. OPTIMAL BANDWIDTH LIMITING

In this section, we investigate the existence of an optimal BL point, i.e., optimal values of the fractions $\Omega_k^{(i,\text{BL})}$, with $i \in \{1,2\}$ such that $\Phi_{\infty}^{(i),\text{BL}} \geqslant \Phi_{\infty}^{(i)}$.

*3) Scenario 1: Spectral Resource Partition:* To improve the NSE by introducing BL in the network, the following condition must be met,

$$\begin{aligned}
\Phi_{\infty}^{(1)} &\leqslant \Phi_{\infty}^{(1),\text{BL}} \\
\frac{1 - \left(\Omega_{\infty}^{(1)}\right)^K}{1 - \Omega_{\infty}^{(1)}} \bar{R}_{\infty} &\leqslant \frac{KL^*}{N} \bar{R}_{\infty} \\
L^* &\geqslant \frac{N}{K} \frac{1 - \left(\Omega_{\infty}^{(1)}\right)^K}{1 - \Omega_{\infty}^{(1)}} \qquad (30)
\end{aligned}$$

In expression above we show that the optimal BL parameter $L^*$ depends mainly on the network load (transmitters per channel, $\frac{K}{N}$) and the SNR of the transmitters. Note that the factor $\Omega_{\infty}^{(1)}$ is a function of $p_{\max}$, $\sigma^2$ and the probability distribution of the channels gains $f_g(\lambda)$ described in (Eq. 15).

*4) Scenario 2: Spectral Resource Sharing:* Following the same reasoning as in scenario 1, we improve the NSE by using BL, if

$$\Phi_{k,\infty}^{(2)} \leqslant \Phi_{k,\infty}^{(2),\text{BL}}. \qquad (31)$$

However, under BL we have that $\Omega_{k,\infty}^{(2),\text{BL}} \leqslant \Omega_{k,\infty}^{(2)}$ and thus,

$$\sum_{k=1}^{K} \Omega_{k,\infty}^{(2)} \bar{R}_{k,\infty}^{(2)} \geqslant \sum_{k=1}^{K} \Omega_{k,\infty}^{(2),\text{BL}} \bar{R}_{k,\infty}^{(2),\text{BL}}. \qquad (32)$$

Then, we have shown that in the asymptotic regime, any kind of BL does not bring any improvement on the NSE. On the contrary, it might introduce significant losses of NSE.

## VI. SIMULATION RESULTS

In this section, we provide numerical results of our mathematical model. First, we compare the asymtotical expressions of the NSE with those obtained by simulations, for both scenarios. In Fig. 2, we plot the NSE of a network with 2 transmitters. Therein, we observe that our asymptotic model perfectly describes the system in the finite case i.e., when $K$ and $N$ are small numbers.

We present also simulations of the NSE obtained in both scenarios as a function of the BL parameter $L$ for different network loads. In Fig. 3, we observe the existence of an optimum BL point for scenario 1. Conversely, in the second scenario, the existence of such optimal is not evident, as at a certain point, the NSE is invariant with respect of the BL parameter $L$.

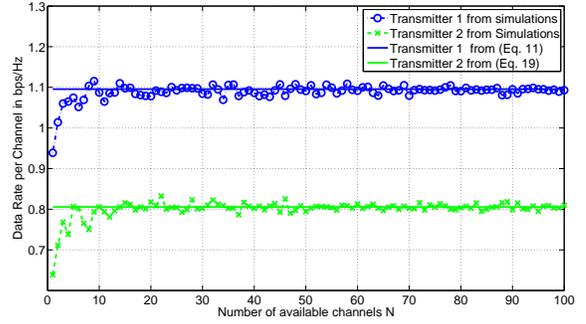

Fig. 2. Data rate per channel in bps/Hz as a function of the number of available channels in the scenario 2. Dashed lines are obtained from simulations considering $10 \log_{10}(\frac{p_{\max}}{\sigma^2}) = 20 dB$ for transmitter 1 and 2. Straight lines are obtained from expression (Eq. 11) for transmitter $k = 1$ and expression (Eq. 20) for $k = 2$.

We compare the optimal BL parameter $L$ obtained from simulations with that obtained from expression (Eq. 30). In Fig. 4 we plot both results. Therein, we show that the asymptotical approximation (Eq. 30) is a precise approximation of the optimal number of channels each transmitter must use to maximize the NSE.

Finally, we show in Fig. 5 the NSE obtained with absence and presence of BL. In the first scenario, we observe a significant gain in NSE when BL is used. This gain is more important for non-overloaded networks, whereas for quasi full-loaded or overloaded networks ($K \geqslant N$), the gain obtained by BL approaches that of limiting the transmitters to use a unique channel. In the same figure, we observe that the NSE appears to be constant for certain intervals. This is due to the fact that inside those intervals the optimal BL parameter remains constant, as shown in Fig. 4. Moreover, the gain in NSE is very significant at high SNR (SNR $= \frac{p_{\max}}{\sigma^2}$) levels. On the contrary, for low SNR levels, small gains in NSE are obtained when the network is low loaded. In Fig. 6 we plot the NSE for several values of SNR in the second scenario. In any case, we observe that there is not significant gain when all transmitters use the same BL parameter $L$.

## VII. CONCLUSIONS

We have shown, that in a decentralized vector MAC where each transmitter aims to maximize its own data rate by using water-filling based power allocation, the network sum-rate can be improved by limiting the number of available channels for each transmitter (bandwidth limiting). We provide closed form expressions for the optimal maximum number of channels each transmitter must access in the case where transmitters use non-intersecting sets of channels. In this case, such an optimum operating point depends mainly on the network load (transmitters per channel) and the different signal to noise ratios. Contrary to the first scenario, in the case of spectrum resource sharing, we show that when all transmitters use the same BL policy, BL does not bring a significant improvement on the network spectral efficiency. Further studies will focus to the case when transmitters have different channel statistics, since it might lead to the usage of different BL policies for each transmitter.

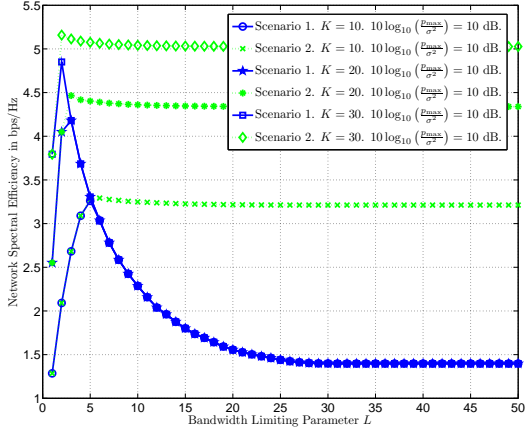

Fig. 3. Network Spectral Efficiency (Eq. 10) in bps/Hz as a function of the maximum number of accessible channels $L$. Total number of available channels $N = 50$, and $10\log_{10}\left(\frac{p_{\max}}{\sigma^2}\right) = 10dB$.

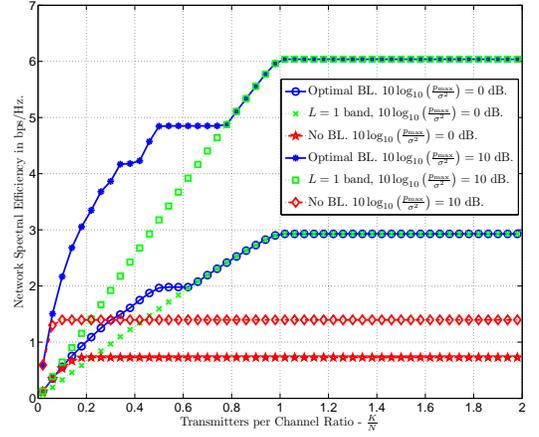

Fig. 5. Network Spectral Efficiency (Eq. 10) in bps/Hz for scenario 1 as a function of the network load $\left(\frac{K}{N}\right)$. Total number of available channels $N = 50$.

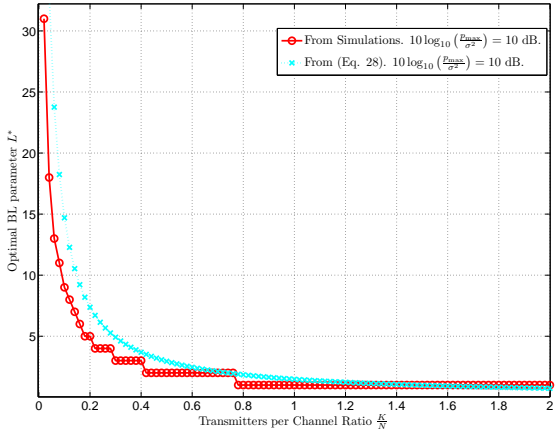

Fig. 4. Optimal BL parameter $L$ (Eq. 30) for scenario 1 as a function of the network load, $\left(\frac{K}{N}\right)$. Total number of available channels $N = 50$, and $10\log_{10}\left(\frac{p_{\max}}{\sigma^2}\right) = 10dB$.

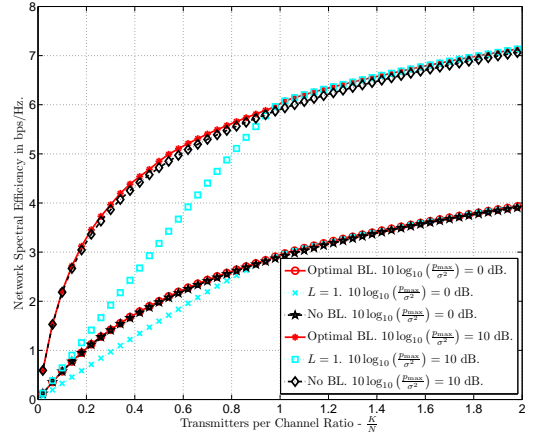

Fig. 6. Network Spectral Efficiency (Eq. 10) in bps/Hz for scenario 2 as a function of the network load $\left(\frac{K}{N}\right)$. Total number of available channels $N = 50$.

## VIII. ACKNOWLEDGMENT

This work is supported by the European Commission in the framework of the FP7 Network of Excellence in Wireless Communications NEWCOM++.